\documentclass[letter]{jpsj2} 
\title{Triplet Pairing Superconductivity Induced by\\ Short-Range Ferromagnetic Correlations in Sr$_2$RuO$_4$}

\author{Kengo {\sc Hoshihara} and Kazumasa {\sc Miyake}}

\inst{Division of Materials Physics, Department of Materials Engineering Science,\\
Graduate School of Engineering Science, Osaka University, Toyonaka 560-8531}

\recdate{\today}

\abst{The microscopic origin of triplet superconductivity in Sr$_2$RuO$_4$ is discussed, paying attention to the role of Coulomb interaction, $U_{pp}$, at the O site. It is shown on the $d$-$p$ model that $U_{pp}$ induces a ferromagnetic exchange interaction between "d-electrons" (molecular orbital with $d_{xy}$-symmetry) at adjacent Ru sites, leading to short-range ferromagnetic correlations and promoting Cooper pairing with ($\sin p_x\pm\ii\sin p_y$)-symmetry on the $\gamma$-band. The reason why such ferromagnetic correlations work effectively may be traced back to the fact that the level of $4d$-electrons at Ru sites is relatively low and located near that of $2p$-electrons at O sites.}

\kword{Sr$_2$RuO$_4$, superconductivity, ferromagnetic correlations, second-order perturbation}

\begin{document}
\newcommand{\non}{\nonumber}
\newcommand{\ii}{{\rm i}}
\maketitle

Since its discovery by Maeno {\it et al.}~\cite{Maeno1}, superconductivity in Sr$_2$RuO$_4$ has attracted much attention both experimentally and theoretically.  Now it seems to have been established experimentally that the superconducting (SC) state is in the chiral spin triplet state~\cite{Muon,Ishida1}. The existence of technically linelike nodes in the SC gap has also been suggested~\cite{Maeno2,Ishida2,Mukuda}. Since these observations are in contrast with the expectation at the early stage of theoretical works~\cite{Rice,Agterberg}, some phenomenological gap models have been proposed to explain those experiments~\cite{MN,Hasegawa,Zhitomirsky}. The results of ref.~\citen{MN} were consistent with the observed gap with linelike nodes in [100] and [010] directions~\cite{Deguchi}. On the other hand, microscopic calculations have been performed on various models and methods~\cite{Takimoto,Ogata,Nomura,Kuroki1}. The results on the multiband Hubbard model by Nomura and Yamada~\cite{Nomura} seem consistent with almost all the available measurements to date probing the gap structure. However, the Hubbard model with only on-site repulsion, $U_{dd}$, seems too stoical to discuss the superconductivity in Sr$_2$RuO$_4$ because the quasi-particles (QP) consist of $4d$-electrons at Ru sites and $2p$-ones at O sites. Then the intersite interaction between QP, arising through the Coulomb interaction at O sites, may not be neglected, while the direct Coulomb interaction $V$ between $4d$-electrons at the nearest-neighbor Ru sites~\cite{Arita} would be negligibly small. The interaction overlooked so far is
\begin{equation}
\mathcal{H}_{\rm ex}=-{U_{pp}\over 6}\!\sum_{m,i}\sum_{\alpha\beta\gamma\delta}p_{mi\alpha}^{\dagger}p_{mi\gamma}^{\dagger}p_{mi\delta}p_{mi\beta}({\vec \sigma}_{\alpha\beta}\!\cdot\!{\vec \sigma}_{\gamma\delta}), \label{eq:1.1}
\end{equation}
where $p_{mi\sigma}$ is the operator for a 2$p_m$-electron ($m=x,y$) of spin $\sigma$ at the $i$th site, and ${\vec \sigma}_{\alpha\beta}$'s are the Pauli matrices. Such a Coulomb interaction gives rise to an effective intersite correlation $J$ between the molecular orbital with $d_{xy}$-symmetry ("d-electron"). We expect that eq.~(\ref{eq:1.1}) can be expressed phenomenologically as
\begin{equation}
\mathcal{H}_{\rm ex}\simeq-\!\!\sum_{\mathbf{k}\mathbf{k'}\mathbf{q}}\sum_{\alpha\beta\gamma\delta}\!J_{\mathbf{q}}\widetilde{d}_{\mathbf{k+q}\alpha}^{\dagger}\widetilde{d}_{\mathbf{k'-q}\gamma}^{\dagger}\widetilde{d}_{\mathbf{k'}\delta}\widetilde{d}_{\mathbf{k}\beta}({\vec \sigma}_{\alpha\beta}\!\cdot\!{\vec \sigma}_{\gamma\delta}), \label{(2)}
\end{equation}
where $\widetilde{d}$ denotes the operator for a "d-electron" and $J_{\mathbf{q}}=2J(\cos q_x+\cos q_y)$. Although exchange interaction (\ref{(2)}) is a too simplified version, we derive a more realistic and complicated version later starting with the $d$-$p$ model and show that this kind of intersite interaction promotes short-range ferromagnetic correlations (SRFMCs), which have been measured quite recently by inelastic neutron scattering~\cite{Braden}. The purpose of this Letter is to demonstrate that SRFMC promotes triplet pairing with ($\sin p_x\pm\ii\sin p_y$)-symmetry, as proposed in ref.~\citen{MN}. We follow the weak-coupling calculations with the pairing interaction given by the second-order perturbation theory (SOPT) with respect to $U_{dd}$ and $U_{pp}$.

\begin{figure}[!h]
\begin{center}
\includegraphics[width=9cm]{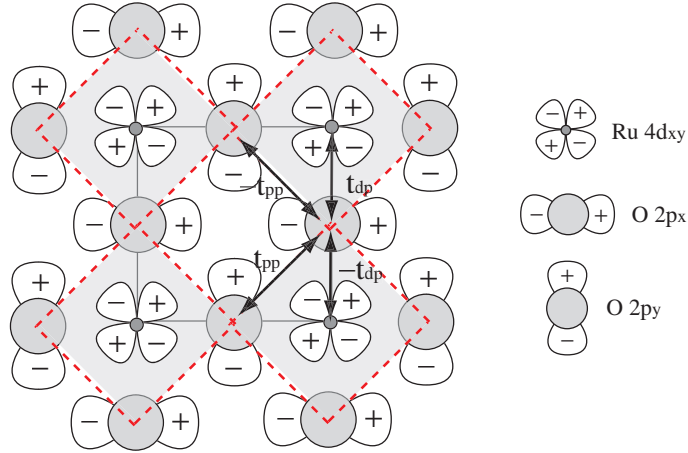}
\caption{Schematic figure of RuO$_2$ plane. $t_{dp}$ and $t_{pp}$ are hopping integrals. Gray domains circled by dashed squares represent the molecular orbitals, which form a two-dimensional square lattice.}\label{RuO-plane}
\end{center}
\end{figure}

Let us consider the situation drawn in Fig.~\ref{RuO-plane}, where a 2$p_x$- or 2$p_y$-orbital is sandwiched between two $d$-orbitals with $d_{xy}$-symmetry.
Hereafter, we restrict our discussion within the $\gamma$-band since it is expected to provide the main contribution~\cite{Nomura}. The noninteracting $d$-$p$ Hamiltonian, which can well describe the situation shown in Fig.~\ref{RuO-plane}, is given as follows:
\begin{equation}
{\mathcal H}^{(0)}_{dp}=\sum_{\mathbf{k},\sigma}\left(d^\dagger_{\mathbf{k}\sigma}\ p^\dagger_{x\mathbf{k}\sigma}\ p^\dagger_{y\mathbf{k}\sigma}\right)
 \left(\begin{array}{ccc}
 \varepsilon_d & V^\ast_{y\mathbf{k}} & V^\ast_{x\mathbf{k}} \\
 V_{y\mathbf{k}} & \varepsilon_p & W_{\mathbf{k}} \\
 V_{x\mathbf{k}} & W_{\mathbf{k}} & \varepsilon_p
 \end{array}\right)
 \left(\begin{array}{c}
 d_{\mathbf{k}\sigma} \\
 p_{x\mathbf{k}\sigma} \\
 p_{y\mathbf{k}\sigma}
 \end{array}\right)
\equiv\sum_{\mathbf{k},\sigma}\vec{\phi}^\dagger_{\mathbf{k}\sigma}\hat{\mathcal{H}}_{\mathbf{k}}^{(0)}\vec{\phi}_{\mathbf{k}\sigma},
\end{equation}
where $d_{\mathbf{k}\sigma}$ and $p_{m\mathbf{k}\sigma}$ are the operators for 4$d$- and 2$p_m$-electrons of momentum $\mathbf{k}$ and spin $\sigma$, respectively, $V_{m\mathbf{k}}=2\ii t_{dp}\sin({k_m\over2})$, and $W_{\mathbf{k}}=4t_{pp}\sin({k_x\over2})\sin({k_y\over2})$. Note that we should take $t_{pp}<0$ for discussing Sr$_2$RuO$_4$, while $t_{pp}>0$ for the cuprates. Solving the equation det$(\hat{\mathcal{H}}_{\mathbf{k}}^{(0)}-\lambda_{\mathbf{k}}\hat{1})=0$, we can obtain the eigenvalue $\lambda_{i\mathbf{k}}$ and eigenvector $\vec{u}_{i\mathbf{k}}$ $(i=1,2,3)$. Then, the diagonalized Hamiltonian becomes
\begin{equation}
{\mathcal H}^{(0)}_{dp}=\sum_{\mathbf{k},\sigma}\vec{\psi}^\dagger_{\mathbf{k}\sigma}\hat{\mathcal{H}}_{\mathbf{k}}^{D}\vec{\psi}_{\mathbf{k}\sigma}.
\end{equation}
Here, $\vec{\phi}_{\mathbf{k}\sigma}$ is related to $\vec{\psi}_{\mathbf{k}\sigma}$ by the relation $\vec{\phi}_{\mathbf{k}\sigma}=\hat{U}^{-1}_{\mathbf{k}}\vec{\psi}_{\mathbf{k}\sigma}$, where the matrix $\hat{U}_{\mathbf{k}}$ is defined as $\hat{U}_{\mathbf{k}}=\left(\vec{u}_{1\mathbf{k}}\ \vec{u}_{2\mathbf{k}}\ \vec{u}_{3\mathbf{k}}\right)$. Substituting the relations between $\vec{\phi}_{\mathbf{k}\sigma}$ and $\vec{\psi}_{\mathbf{k}\sigma}$ into eq.~(\ref{eq:1.1}), we obtain, as shown below, the effective interaction between QP described by $a_{i\mathbf{k}\sigma}$. For this purpose, it is enough to confine our discussions within the case $t_{pp}=0$: namely we need not consider the hopping $t_{pp}$. Because electrons located on both sides of a certain O site can interact with each other effectively only through the O site, the term including $t_{pp}$ corresponds to the next nearest-neighbor interaction (higher-order contribution). Then, hereafter, we can use simplified relations such as
\begin{eqnarray}
d_{\mathbf{k}\sigma}={\lambda_{1\mathbf{k}}\over L_{1\mathbf{k}}}a_{1\mathbf{k}\sigma}+{\lambda_{2\mathbf{k}}\over L_{2\mathbf{k}}}a_{2\mathbf{k}\sigma}+{\lambda_{3\mathbf{k}}\over L_{3\mathbf{k}}}a_{3\mathbf{k}\sigma}, \label{d}
\end{eqnarray}
\begin{eqnarray}
p_{m\mathbf{k}\sigma}={V_{\overline{m}\mathbf{k}}\over L_{1\mathbf{k}}}a_{1\mathbf{k}\sigma}+{V_{\overline{m}\mathbf{k}}\over L_{2\mathbf{k}}}a_{2\mathbf{k}\sigma}+{V_{\overline{m}\mathbf{k}}\over L_{3\mathbf{k}}}a_{3\mathbf{k}\sigma}, \label{px}
\end{eqnarray}
where $L_{i\mathbf{k}}=(|V_{x\mathbf{k}}|^2+|V_{y\mathbf{k}}|^2+\lambda_{i\mathbf{k}}^2)^{1/2}$ ($i=$1, 2, 3). Moreover, it is sufficient to consider only the QP on the orbital $i=1$, because the other bands are completely filled with electrons. Hereafter, we denote $a_{1\mathbf{k}\sigma}$ and $L_{1\mathbf{k}}$ simply as $a_{\mathbf{k}\sigma}$ and $L_{\mathbf{k}}$, respectively. The operator $a_{\mathbf{k}\sigma}$ is regarded as the $\widetilde{d}_{\mathbf{k}\sigma}$ in eq.~(\ref{(2)}).

Now we introduce the interaction terms. Using relations (\ref{d}) and (\ref{px}), we express such terms using $a_{\mathbf{k}\sigma}$'s. Substituting relation (\ref{px}), into the Fourier transformed form of eq.~(\ref{eq:1.1}), we obtain
\begin{eqnarray}
{\mathcal H}^{(x)}_{\rm ex}=-\frac{U_{pp}}{6}\sum_{\mathbf{kk'q}}\sum_{\alpha\beta\gamma\delta}{V_{x\mathbf{k+q}}V_{x\mathbf{k'-q}}V_{x\mathbf{k'}}V_{x\mathbf{k}}\over L_{\mathbf{k+q}}L_{\mathbf{k'-q}}L_{\mathbf{k'}}L_{\mathbf{k}}}a^{\dagger}_{\mathbf{k+q}\alpha}a^{\dagger}_{\mathbf{k'-q}\gamma}a_{\mathbf{k'}\delta}a_{\mathbf{k}\beta}(\vec{\sigma}_{\alpha\beta}\!\cdot\!\vec{\sigma}_{\gamma\delta}). \label{H}
\end{eqnarray}
At first sight, it seems that this interaction Hamiltonian, eq.~(\ref{H}), has a complicated form in real space. However, if we neglect the $\mathbf{k}$-dependence of $L_{\mathbf{k}}$'s where $L_{\mathbf{k}}\equiv\Delta\sim O(\varepsilon_d-\varepsilon_p)$, we can perform the Fourier transformation of eq.~(\ref{H}), leading to
\begin{eqnarray}
 & & \hspace{-1.2cm}\mathcal{H}_{\rm ex}=\frac{-U_{pp}t_{dp}^4}{3\Delta^4}\sum_{\langle i,j\rangle}\sum_{\alpha\beta\gamma\delta}\bigl\{(i,i,i,i)+(i,j,j,i) \non \\
 & & \mbox{}+(i,j,i,j)+(i,i,j,j)-(i,i,i,j)-(i,i,j,i) \non \\
 & & \mbox{}-(j,i,i,i)-(i,j,i,i)\bigr\}(\vec{\sigma}_{\alpha\beta}\!\cdot\!\vec{\sigma}_{\gamma\delta}), \label{sss}
\end{eqnarray}
where we have added the term arising from the exchange process along the $y$-direction, the sum $\langle i,j\rangle$ is taken within the nearest-neighbor sites, and the abbreviation $(i,j,k,l)$ represents the term $a^{\dagger}_{i\alpha}a^{\dagger}_{j\gamma}a_{k\delta}a_{l\beta}$. Eventually, it is revealed that eq.~(\ref{H}) contains all the types of interaction between two electrons at nearest-neighbor Ru sites. Performing the Fourier transformation again and executing straightforward calculations, we can rewrite eq.~(\ref{sss}) into the following form:
\begin{eqnarray}
\mathcal{H}_{\rm ex}=-{1\over4}\sum_{\mathbf{kk'q}}\sum_{\alpha\beta\gamma\delta}J_{\mathbf{k},\mathbf{k'};\mathbf{q}}a^{\dagger}_{\mathbf{k+q}\alpha}a^{\dagger}_{\mathbf{k'-q}\gamma}a_{\mathbf{k'}\delta}a_{\mathbf{k}\beta}(\vec{\sigma}_{\alpha\beta}\!\cdot\!\vec{\sigma}_{\gamma\delta}-\delta_{\alpha\beta}\delta_{\gamma\delta}), \label{ex}
\end{eqnarray}
\begin{eqnarray}
J_{\mathbf{k},\mathbf{k'};\mathbf{q}}=\frac{2U_{pp}t_{dp}^4}{L_{\mathbf{k+q}}L_{\mathbf{k'-q}}L_{\mathbf{k'}}L_{\mathbf{k}}}\sum_m\Bigl\{{1\over2}+\cos q_m-\cos k_m-\cos k_m'+{1\over2}\cos(k_m+k_m')\Bigr\}. \label{12}
\end{eqnarray}
In eq.~(\ref{12}), the $\mathbf{k}$-dependence of $L_{\mathbf{k}}$'s is recovered again. The second term in the curly brackets of eq.~(\ref{12}), $\cos q_m$, corresponds to the simplified version of eq.~(\ref{(2)}). There are many terms which are not included in the naive estimation, eq.~(\ref{(2)}). Therefore, the coupling constant $J$ cannot be written simply as $J_{\mathbf{q}}$, but has a complicated momentum dependence $J_{\mathbf{k},\mathbf{k'};\mathbf{q}}$ as eq.~(\ref{12}). The term $\delta_{\alpha\beta}\delta_{\gamma\delta}$ in eq.~(\ref{ex}), which appears automatically through the procedure above, assures that, due to the Pauli principle, there is no interaction between electrons with the parallel spin direction originally on the same O site.
\begin{figure}[!h]
\begin{center}
\includegraphics[width=9cm]{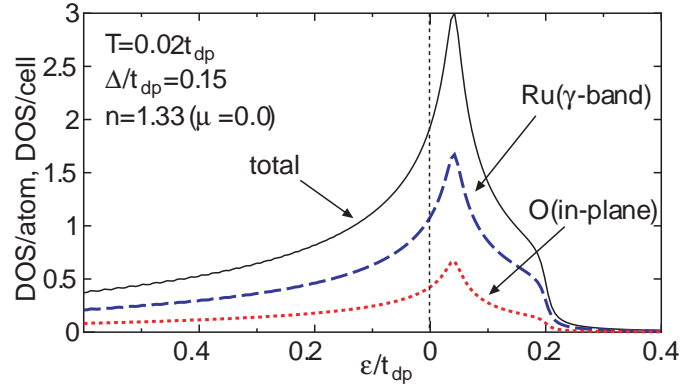}
\caption{Total and partial DOS in RuO$_2$ plane.}\label{zu-DOS}
\end{center}
\end{figure}
For interaction (\ref{ex}) to play an important role, the ratio $t_{dp}/(\varepsilon_d-\varepsilon_p)\equiv t_{dp}/\Delta$ should not be too small (namely, $\Delta/t_{dp}$ should not be too large). Such a condition is expected to be fulfilled in Ru-oxides because the level of $4d$-electrons in Ru is deeper than that of $3d$-electrons in Fe or of $5d$-electrons in Os in general, and the level is located near that of $2p$-electrons at the O site. This implies that there exists a strong hybridization between Ru and O. Indeed, as shown in Fig.~\ref{zu-DOS}, the partial density of states (DOS) for $4d_{xy}$, $2p_{x}$ and $2p_{y}$ obtained by band structure calculations~\cite{Oguchi,Singh} are reproduced by taking $\Delta/t_{dp}\simeq0.15$ with the use of eqs.~(\ref{d}) and (\ref{px}): namely, $\rho_{\gamma}\simeq2.0\rho_{\rm O}$ ($\rho_{\gamma}\simeq0.5\rho_{\rm Ru}$) where $\rho_{\gamma}$, $\rho_{\rm Ru}$ and $\rho_{\rm O}$ denote the partial DOS per atom (at the Fermi level) of Ru ($\gamma$-band), Ru (total of $\alpha$-, $\beta$-, and $\gamma$-bands) and O (in-plane), respectively. The same procedure is applicable to the on-site Coulomb repulsion $U_{dd}$ between $d$-electrons, $U_{dd}\sum_id^{\dagger}_{i\uparrow}d_{i\uparrow}d^{\dagger}_{i\downarrow}d_{i\downarrow}$. Along the way parallel to obtaining exchange interaction (\ref{H}) from the original interaction between $p$-electrons, eq.~(\ref{eq:1.1}), the Hubbard interaction is expressed in terms of the $a_{\mathbf{k}\sigma}$ operators describing the QP band. Then, the total Hamiltonian is given by
\begin{equation}
\mathcal{H}=\sum_{\mathbf{k},\sigma}(\lambda_{\mathbf{k}}-\mu)a_{\mathbf{k}\sigma}^{\dagger}a_{\mathbf{k}\sigma}+\sum_{\mathbf{kk'q}}U_{\mathbf{k},\mathbf{k'};\mathbf{q}}a^{\dagger}_{\mathbf{k+q}\uparrow}a^{\dagger}_{\mathbf{k'-q}\downarrow}a_{\mathbf{k'}\downarrow}a_{\mathbf{k}\uparrow}+\mathcal{H}_{\rm ex}, \label{fullH}
\end{equation}
\begin{equation}
U_{\mathbf{k},\mathbf{k'};\mathbf{q}}={U_{dd}\lambda_{\mathbf{k+q}}\lambda_{\mathbf{k'-q}}\lambda_{\mathbf{k'}}\lambda_{\mathbf{k}}\over 8L_{\mathbf{k+q}}L_{\mathbf{k'-q}}L_{\mathbf{k'}}L_{\mathbf{k}}}, \label{Ukkk}
\end{equation}
where $\mu$ denotes the chemical potential. It is noted that the repulsion $U_{dd}$ in eq.~(\ref{Ukkk}) should be regarded as an effective interaction between QP, that is reduced by the electron correlation leaving a relatively heavy QP. We set $t_{pp}=-0.4t_{dp}$ to reproduce the Fermi surface of the $\gamma$-band given by the band structure calculations. Then the bandwidth of the $\gamma$-band is $W\sim2.2t_{dp}$. The last two terms of eq.~(\ref{fullH}) can be rewritten into a more convenient form to perform the perturbation expansion as follows:
\begin{equation}
\mathcal{H}_{\rm int}=\sum_{\mathbf{kk'q}}\widetilde{J}_{\mathbf{k},\mathbf{k'};\mathbf{q}}a^{\dagger}_{\mathbf{k+q}\uparrow}a^{\dagger}_{\mathbf{k'-q}\downarrow}a_{\mathbf{k'}\downarrow}a_{\mathbf{k}\uparrow},
\end{equation}
\begin{equation}
\widetilde{J}_{\mathbf{k},\mathbf{k'};\mathbf{q}}=U_{\mathbf{k},\mathbf{k'};\mathbf{q}}+J_{\mathbf{k},\mathbf{k'};\mathbf{q}}+J_{\mathbf{k'},\mathbf{k};\mathbf{k-k'+q}}.
\end{equation}
Hereafter, $t_{dp}$ is adopted as a unit of energy.

\begin{figure}[!h]
\begin{center}
\includegraphics[width=10cm]{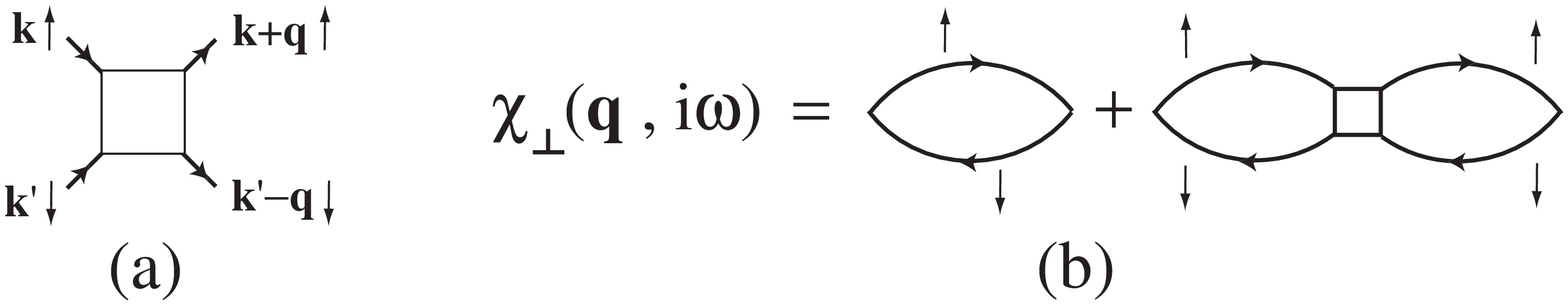}
\caption{(a) Vertex $\widetilde{J}_{\mathbf{k},\mathbf{k'};\mathbf{q}}$. (b) Spin susceptibility $\chi_{\perp}(\mathbf{q},\ii\omega)$.}\label{dia-chi}
\end{center}
\end{figure}

\begin{figure}[!h]
\begin{center}
\includegraphics[width=9cm]{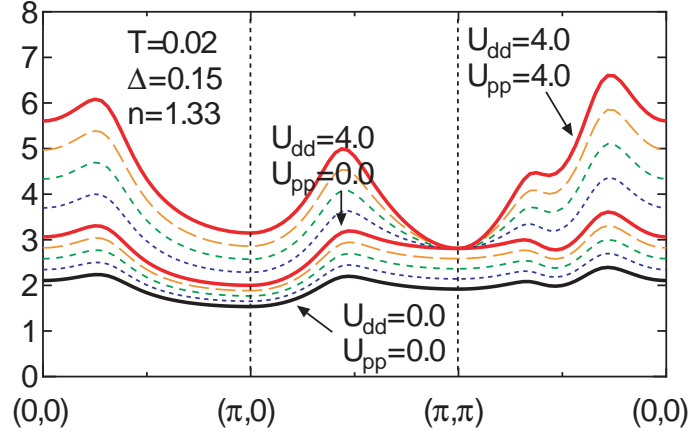}
\caption{Momentum dependence of static spin susceptibility $\chi_{\perp}(\mathbf{q},0)$ for several values of $U_{dd}$ and $U_{pp}$. The dashed lines that have no parameter values correspond to, from the lower line, $(U_{dd},U_{pp})=(1,0), (2,0)$ and so on, respectively. Namely, we first applied $U_{dd}$, then applied $U_{pp}$.}\label{zu-chi}
\end{center}
\end{figure}

The static spin susceptibility $\chi_{\perp}(\mathbf{k},0)$ including vertex correction, up to the first-order perturbation, is given by
\begin{equation}
\chi_{\perp}(\mathbf{q},\ii\omega)=-T\!\sum_{\varepsilon,\mathbf{p}}\Bigl\{1-T\!\sum_{\varepsilon',\mathbf{p'}}\widetilde{J}_{\mathbf{p+q},\mathbf{p'};\mathbf{-p+p'}}\mathcal{G}(\mathbf{p'+q},\ii\varepsilon'+\ii\omega)\mathcal{G}(\mathbf{p'},\ii\varepsilon)\Bigr\}\mathcal{G}(\mathbf{p+q},\ii\varepsilon+\ii\omega)\mathcal{G}(\mathbf{p},\ii\varepsilon),
\end{equation}
whose diagrammatic expression is shown in Fig.~\ref{dia-chi}(b). It is enough to calculate $\chi_{\perp}$ because the SU(2) symmetry is preserved.
The results of numerical calculations are shown in Fig.~\ref{zu-chi}. $\chi_{\perp}(\mathbf{q},0)$ has some peaks at some incommensurate wave vectors $\mathbf{Q}_{\rm ic}$'s. The Fermi surface simulating the $\gamma$-band of Sr$_2$RuO$_4$ has no nesting tendency, although the Fermi surface shifted by $\mathbf{Q}_{\rm ic}$'s has a point contact in the extended Brillouin zone scheme~\cite{Bergemann}. Therefore, the peak structure of $\chi_{\perp}(\mathbf{q},0)$ at $\mathbf{Q}_{\rm ic}$'s is considered to come from the band effect. In particular, the shape of $\chi_{\perp}(\mathbf{q},0)$ consists of a broad peak around $\mathbf{q}=(0,0)$ and a small dip (at $\mathbf{q}\sim(0,0)$), which gives rise to a practical SRFMC. The effect of $U_{dd}$ is to enhance $\chi_{\perp}(\mathbf{q},0)$ in the entire $\mathbf{q}$-space as seen in Fig.~\ref{zu-chi}. On the other hand, the peak around $\mathbf{q}=(0,0)$ is much more enhanced by $U_{pp}$ than by $U_{dd}$, while there is no extra increase by $U_{pp}$ at $(\pi,\pi)$. The qualitative behavior of $\chi_{\perp}(\mathbf{q},0)$ remains unchanged if we neglect the $\mathbf{k}$-dependence of $L_{\mathbf{k}}$'s in eq.~(\ref{12}) or (\ref{Ukkk}). Then, we can draw the picture that SRFMC is induced by the effect of on-site repulsion between $p$-electrons at the O site.
The broad peak structure at $\mathbf{q}=(0,0)$ is consistent with the SRFMC that has been measured quite recently by inelastic neutron scattering~\cite{Braden}, and is responsible for the occurrence of triplet superconductivity in Sr$_2$RuO$_4$ as discussed below. As for the results of different electron fillings, the more electrons are doped, the more the peak structure around $\mathbf{q}=(0,0)$ is enhanced.

\begin{figure}[!h]
\begin{center}
\includegraphics[width=12cm]{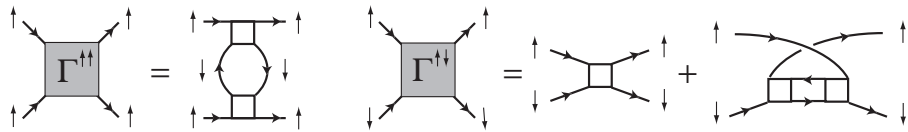}
\caption{Pairing interaction vertices $\Gamma^{\uparrow\uparrow}$ and $\Gamma^{\uparrow\downarrow}$ up to the second-order perturbation with respect to $\widetilde{J}$. The solid line and the square denote the noninteracting Matsubara Green function $\mathcal{G}$ and the intersite coupling vertex $\widetilde{J}$, respectively.}\label{pair}
\end{center}
\end{figure}

Diagrams for the irreducible pairing vertices up to SOPT in terms of $\widetilde{J}$ are shown in Fig.~\ref{pair}.
Its analytic expressions are given as follows:
\begin{eqnarray}
\Gamma^{\uparrow\uparrow}_{\mathbf{k,k'}}=T\!\sum_{\varepsilon,\mathbf{p}}\widetilde{J}_{\mathbf{k},\mathbf{p};\mathbf{-k+k'}}\widetilde{J}_{\mathbf{p+k-k'},-\mathbf{k};\mathbf{-k+k'}}\mathcal{G}(\mathbf{p},\ii\varepsilon)\mathcal{G}(\mathbf{p}+\mathbf{k-k'},\ii\varepsilon), \label{pair-t}
\end{eqnarray}
\begin{eqnarray}
\Gamma^{\uparrow\downarrow}_{\mathbf{k,k'}}=\widetilde{J}_{\mathbf{k},\mathbf{-k};\mathbf{-k+k'}}-T\!\sum_{\varepsilon,\mathbf{p}}\widetilde{J}_{\mathbf{k},\mathbf{p};\mathbf{p+k'}}\widetilde{J}_{\mathbf{-k},\mathbf{p+k+k'};\mathbf{p+k}}\mathcal{G}(\mathbf{p},\ii\varepsilon)\mathcal{G}(\mathbf{p}+\mathbf{k+k'},\ii\varepsilon). \label{ud}
\end{eqnarray}
The pairing interaction in the triplet manifold, $V^{\rm t}_{\mathbf{kk'}}$, is given by eq.~(\ref{pair-t}) directly, and that in the singlet manifold, $V^{\rm s}_{\mathbf{kk'}}$, is given by
\begin{eqnarray}
V^{\rm s}_{\mathbf{k,k'}}=\widetilde{J}_{\mathbf{k},\mathbf{-k};\mathbf{-k+k'}}-T\!\sum_{\varepsilon,\mathbf{p}}\widetilde{J}_{\mathbf{k},\mathbf{p};\mathbf{p-k'}}\widetilde{J}_{\mathbf{-k},\mathbf{p+k-k'};\mathbf{p+k}}\mathcal{G}(\mathbf{p},\ii\varepsilon)\mathcal{G}(\mathbf{p}+\mathbf{k-k'},\ii\varepsilon). \label{pair-s}
\end{eqnarray}
Note that SU(2) symmetry is preserved in the triplet manifold. Namely, the triplet component stemming from eq.~(\ref{ud}) is the same as the one from eq.~(\ref{pair-t}).

\begin{figure}[!h]
\begin{center}
\includegraphics[width=9cm]{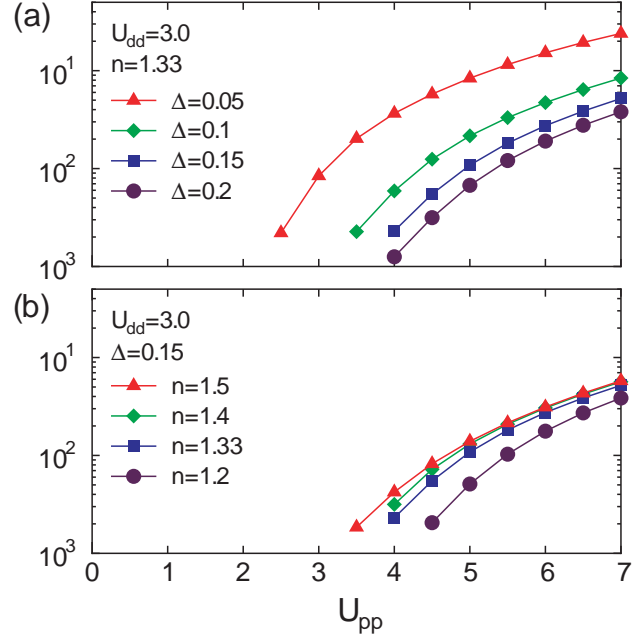}
\caption{$T_c$ for triplet pairing with $p_x$-symmetry as a function of $U_{pp}$. (a) $\Delta$ dependence. (b) doping dependence.}\label{zu-Tc}
\end{center}
\end{figure}

At $T=T_c$, we can use the linearized gap equation, in the weak-coupling formalism, given by

\begin{equation}
\Delta_{\mathbf{k}}=-\sum_{\mathbf{k'}}V_{\mathbf{k,k'}}^{s,t}\frac{\Delta_{\mathbf{k'}}}{\xi_{\mathbf{k'}}}\tanh\biggl(\frac{\xi_{\mathbf{k'}}}{2T_c}\biggr). \label{gap-eq}
\end{equation}
A proper way of obtaining the gap structure with the highest $T_c$ is to solve eq.~(\ref{gap-eq}) as an eigenvalue problem without specifying the type of the gap $\Delta_{\mathbf{k}}$. However, for simplicity, we seek the variational solution for eq.~(\ref{gap-eq}). Namely, we calculate $T_c$ by specifying the type of gap function, such as $\sqrt{2}\sin k_x$ ($p_x$-pairing), $\sqrt{2}\sin(k_x+k_y)$ ($p_{x+y}$-pairing) for triplet pairing, and $2\sin k_x\sin k_y$ ($d_{xy}$-pairing), $[\cos k_x-\cos k_y]$ ($d_{x^2-y^2}$-pairing) for singlet pairing. However, it is rational to consider that the type of pairing giving the highest $T_c$ is dominant in the true gap function. The pairing interaction $V^{\rm t}_{\mathbf{k,k'}}$, eq.~(\ref{pair-t}), and $V^{\rm s}_{\mathbf{k,k'}}$, eq.~(\ref{pair-s}) is estimated at $T=0.02$, because their values are not sensitive at temperatures $T<0.02$. We adopt the first Brillouin zone divided into $45\!\times\!45$ $\mathbf{k}$-meshes. The results for a series of parameters are shown in Fig.~\ref{zu-Tc}. We can see in Fig.~\ref{zu-Tc} that $T_c$ is enhanced as $U_{pp}$ is applied. For a system with a fillings of $n=1.33$, a small value of $\Delta$ (splitting between $p$- and $d$-levels), and a moderate value of $U_{pp}$, corresponding to Sr$_2$RuO$_4$, $p_x$-pairing would be realized as shown in Fig.~\ref{zu-Tc}(a). When the system is located away from the half-filling state, it is also enhanced as shown in Fig.~\ref{zu-Tc}(b). On the other hand, $p_{x+y}$-pairing has $T_c$ less than $10^{-3}$.
In ref.~\citen{Arita}, it is discussed that the next nearest-neighbor (diagonal) pairings, $d_{xy}$- or $p_{x+y}$-pairing, are promoted due to the nearest-neighbor Coulomb repulsion $V$. In our case, on the contrary, the nearest-neighbor pairing, $p_x$-pairing, is realized through SRFMC of $\mathcal{H}_{\rm ex}$. As for the other symmetries, $d_{x^2-y^2}$ and $d_{xy}$, we cannot obtain a finite $T_c$ in the parameter region shown in Fig.~\ref{zu-Tc}. At first sight, it appears unrealistic that $U_{pp}$ is larger than $U_{dd}\sim W$, because $3d$-electrons are much more localized on the ion site than $2p$-electrons. However, it is crucial to note that the Coulomb interactions here are effective ones that are greatly renormalized. Indeed, the effect of $U_{pp}$ on QP cannot be screened by avoiding $U_{pp}$ because QP consisting of molecular orbitals at the nearest-neighbor sites cannot get rid of $U_{pp}$. On the other hand, the effective interaction $U_{dd}$ can be reduced by correlations and by making a heavy QP. Therefore the values adopted for $U_{dd}$ and $U_{pp}$ are considered to be not unrealistic.
 
In conclusion, we have derived from the so-called $d$-$p$ model the effective intersite exchange interaction $\mathcal{H}_{\rm ex}$, whose origin is the on-site Coulomb interaction at an O site. SRFMC were induced by this exchange interaction. Then, we have shown that the triplet SC state of ($\sin p_x\pm\ii\sin p_y$)-symmetry is promoted as applying $U_{pp}$ within SOPT. These results simulate quite well the properties observed in Sr$_2$RuO$_4$. While our results are derived on SOPT for the pairing interaction, they might make sense at least in describing the qualitative behavior~\cite{Kondo}. It is possible to take into account the effect of $\alpha$- and $\beta$-bands on the same footing as the $\gamma$-band since the exchange interaction arising from the same mechanism also works between QPs in these bands. According to preliminary calculations, the pairing with [$\sin(p_x+p_y)+\ii\sin(p_x-p_y)$]-symmetry could be promoted, which is consistent with a line-node-like gap along the diagonal direction $k_x=\pm k_y$~\cite{Deguchi,Nomura}.

\section*{Acknowledgements}
We have benefited much from conversations with H. Harima, H. Kohno and T. Takimoto. This work is supported by a Grant-in-Aid for Creative Scientific Research (No. 15GS0213), a Grant-in-Aid for Scientific Research (No. 16340103), and the 21st Century COE Program (G18) of the Japan Society for the Promotion of Science.

\end{document}